%% file: paper-rshetal-pversion.tex
\documentclass{aa}
\usepackage{txfonts}
\usepackage{epsfig}
\usepackage{graphics}
\usepackage{url}
\usepackage{epstopdf}
\usepackage{float}
\usepackage{pslatex}
\usepackage{color}
\input{rgb}
\usepackage{graphicx,natbib}
\bibpunct{(}{)}{;}{a}{}{,}
\usepackage{txfonts}
\begin{document}
   \title{Herschel/PACS\thanks{Herschel is an ESA space observatory with science instruments
provided by European-led Principal Investigator consortia and with important
participation from NASA} spectroscopy of trace gases of the stratosphere of Titan
}


   \author{M. Rengel\inst{1}
          \and
          H. Sagawa\inst{1}\thanks{Current address: National Institute of Information and Communications Technology, Japan} \and
          P. Hartogh\inst{1} \and
          E. Lellouch\inst{2}\and
          H. Feuchtgruber\inst{3}\and
          R. Moreno\inst{2}\and
          C. Jarchow\inst{1}\and
          R. Courtin\inst{2}\and
          J. Cernicharo\inst{4}\and
          L. M. Lara\inst{5}
          }

   \institute{Max-Planck-Institut f\"ur Sonnensystemforschung, Max-Planck-Str. 2, 37191 Katlenburg-Lindau, Germany \\
              \email{rengel@mps.mpg.de}
         \and
             LESIA--Observatoire de Paris, CNRS, Universit\'e Paris 6, Universit\'e Paris-Diderot, 5 place Jules Janssen, 92195 Meudon, France
         \and
              Max-Planck-Institut f\"ur extraterrestrische Physik, Giessenbachstrasse, 85748 Garching, Germany
         \and
              Departamento de Astrof\'isica, Centro de Astrobiolog\'ia, CSIC-INTA, Torrej\'on de Ardoz, 28850 Madrid, Spain
         \and
              Instituto de Astrof\'isica de Andaluc\'ia (CSIC), Granada, Spain}


   \date{Received May 22, 2013; accepted November 6, 2013}

  \abstract
   {}
   {We investigate the composition of Titan's stratosphere from new medium-resolution far-infrared observations performed with the full range of Herschel's Photodetector Array Camera and Spectrometer (PACS)  (51--220\,$\mu$m at a resolution $\lambda$/$\Delta\lambda$ ranging from 950 to 5500 depending on wavelength and grating order).}
   {Using PACS, we obtained the spectral emission of several features of the Titan's stratosphere. We used a line-by-line radiative transfer code and the least-squares fitting technique to infer the abundances of the trace constituents.}
   {Numerous spectral features attributable to CH$_4$, CO, HCN, and H$_2$O are present. From the flux density spectrum measured and by a detailed
comparison with synthetic spectra, we constrain the stratospheric abundance of CH$_4$, which is assumed to be constant with altitude, to be  1.29 $\pm$ 0.03\%. Similarly,
we constrain the abundance of CO to be 50 $\pm$ 2\,ppm, and the HCN vertical distribution consistent with an increase from 40\,ppb at $\sim$100\,km to $\sim$4\,ppm at $\sim$200\,km,
which is an altitude region where the HCN signatures are sensitive. Measurements of three H$_2$O rotational lines
confirm the H$_2$O distribution profile
recently obtained with Herschel. Furthermore, we determine the isotopic ratios $^{12}$C/$^{13}$C in CO and HCN to be
124 $\pm$ 58, and 66 $\pm$ 35, respectively.
Comparisons between our results and the values derived with other instruments show that our results are
consistent with the vertical distributions and isotopic ratios in previous studies, except for the HCN distribution obtained with
Cassini/CIRS,
which does not agree with the PACS lines at the 1-$\sigma$ confidence interval.}
   {}
   \maketitle
%

\section{Introduction}

The nitrogen (N$_2$)-dominated atmosphere of Titan, Saturn's largest moon, exhibits a great diversity and complexity of molecules and high organic material abundances. Many species are formed in the upper atmosphere (above 600\,km) by interaction of solar UV photons, energetic
ions/electrons, and cosmic-ray particles, which initiate a complicated photochemical network.
For several of these molecules, the region between the upper atmosphere and their condensation sink
(60--100\,km) is a transition region.
In the stratosphere (40--320\,km), complex chemistry is still in action, and this region is probed mainly with thermal infrared measurements.
Previous studies of Titan's submillimetre and infrared spectra from space (e.g.
the Infrared Space Observatory (ISO) and Cassini/ Composite InfraRed Spectrometer (CIRS)) and ground-based facilities at
different spectral resolutions have revealed many specific aspects of the molecular composition, permitted a
detailed investigation of the photochemical and dynamical processes in Titan's
atmosphere, and confirmed the seasonal behaviour of Titan's atmospheric composition \citep[e.g.][]{roe2004, dekok2007, coustenis2007, teanby2009, ren11, teanby2012}.
New observations at different spectral ranges, higher spectral resolution and
sensitivity, and at different temporal coverage during a Titan's year (29.5 Earth years) offer the opportunity of exploring and helping to advance the study of the abundance distribution (e.g. new abundance constraints and rate of seasonal variations).

The advent of the Herschel Space Observatory \citep{herschel} allowed to determine
the stratosphere abundances of CH$_4$, CO and HCN with the Spectral and Photometric Imaging Receiver (SPIRE)
covering the 194--671 $\mu$m spectral region with an spectral resolution of 0.04\,cm$^{-1}$ \citep{courtin2011},
to detect HNC with the Heterodyne Instrument for the Far-Infrared (HIFI) \citep{moreno11}, and infer the vertical profile
of H$_2$O over the 100--450\,km altitude range by combining data from HIFI and the Photodetector Array Camera and Spectrometer (PACS) \citep{moreno2012}.

In this paper, we report the results from a full grating resolution spectra of Titan at
 51--220\,$\mu$m (5878--1363\,GHz) taken with PACS \citep{PACS} onboard Herschel.
Although it covers a smaller spectral range than Cassini/CIRS and a range similar to that of ISO/LSW ([17-1000] and
[2.4-196]\,$\mu$m, respectively), PACS has a higher spectral resolution than previous instruments (ISO/SWS and LSW have resolving powers of 1000-2500 and 200, respectively, and Cassini/CIRS has a
spectral resolution of 0.5\,cm$^{-1}$).
We report the gases identified
in the spectra, derive their abundances, and compare them with previous studies. A few isotopic lines are identified,
from which we estimate the isotopic ratios $^{12}$C/$^{13}$C in CO and HCN. More possible detections of isotopes are
limited by the sensitivity of the PACS observations
and the weakness of the lines, which can easily be confused with noise.


\section{Observations and data reduction}

In the framework of the program ``Water and Related Chemistry in the Solar System", \citep{hartogh10},
full-range observations of Titan were performed on June 23 and December 15, 2010 (Table\,1).
Observations were carried out in chopped-nodded PACS range spectroscopy modes \citep{PACS}, at Nyquist/Spectral Energy Distribution (SED)
(second and first orders of the grating spectrometer) and at high spectral sampling density (third and first orders)
for the June and December observations, respectively.

\begin{figure*}
\begin{center}
\begin{minipage}[c]{0.73\linewidth}
\epsfig{bb = 0 0 600 750, file=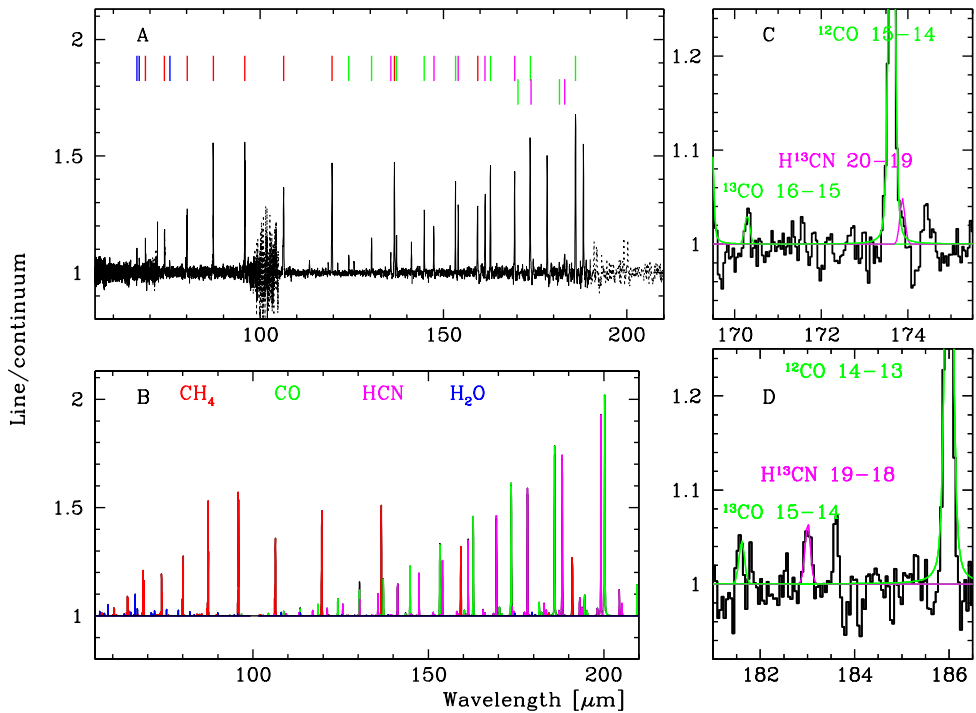, angle=-90, width=\linewidth}
\end{minipage}\hfill
\begin{minipage}[c]{0.2\linewidth}
\caption{\textbf{A}: Black line shows the PACS spectrum of Titan, expressed in line-to-continuum ratios.
The coloured marks at the top show the selected lines and transitions for
the analysis in Sect.\,4. The dashed line indicates spectral regions affected by spectral leakage due to grating order overlap. \textbf{B:} Best-fit synthetic spectra indicating the signatures attributable to CH$_4$, CO, H$_2$O,
and HCN (Sect.\,4). \textbf{C-D}: The [170-176] and [181-187] $\mu$m regions of the PACS spectrum displaying signatures of the isotopologues,
along with the best-fit model with the detected isotopic variants.}
\label{fig:fig1}
\end{minipage}
\end{center}
\end{figure*}

\begin{table}[h]\footnotesize
\caption{Summary of the observations}             
\resizebox{0.48\textwidth}{!}{%
\label{table:1}      
\centering                          
\begin{tabular}{c c c c c}        
\hline\hline                 
Obs. ID & Start time & $T_{obs}$ &  Range$^{a}$ \\
        &  [UTC]     &  [sec.]   &  [$\mu$m]\\
\hline                        
   1342199145 & 23 June 2010 00:42:40 & 1140  & 51--073$^{B2A}$,102--146$^{R1}$\\      
   1342199146 & 23 June 2010 01:03:48 & 2268  & 70--105$^{B2B}$,140--220$^{R1}$\\
   1342211198 & 15  Dec 2010 05:56:53 & 4051  & 55--072$^{B3A}$,165--216$^{R1}$\\
   1342211199 & 15  Dec 2010 07:06:33 & 3509  & 70--099$^{B2B}$,140--198$^{R1}$\\
   1342211200 & 15  Dec 2010 08:07:14 & 1983  & 51--071$^{B2A}$,102--141$^{R1}$\\
\hline                                   
\end{tabular}}
Notes. $^{a}$= spectral band
\end{table}

The entire spectral range of PACS was measured at full instrumental resolution $\lambda$/$\Delta \lambda$, ranging
from 950 to 5500 depending on wavelength and grating order \citep{PACS}. Since blue and red spectrometer data are
acquired in parallel, several spectral ranges were observed in overlap.
To minimise spectral contributions from Saturn, Titan was observed at the largest feasible elongations from Saturn (168$\arcsec$ and 173$\arcsec$ on 23 June and 15 December 2010, respectively).
The data processing from Level\,0 to 1 was carried out by standard PACS pipeline modules using the Herschel Interactive Processing Environment (HIPE v.4 and v.6). All subsequent processing (flat-fielding, outlier removal and rebinning)
was carried out with standard Interactive Data Language (IDL) tools.  Because the absolute calibration accuracy is limited by detector response drifts and slight pointing
offsets\footnote{http://herschel.esac.esa.int/},
the resulting spectrum was
divided by its continuum, a step that is not subject to changes through later HIPE releases.
For spectral ranges covered more than once, the observation with the highest resolution was selected.
When observations are covered with the same spectral band,
an averaged observation was used to increase the S/N. The data adquired during different operational days were carefully compared to verify that no remarkable temporal changes  in the instrumental characterists and in
seasonal changes in the atmosphere of Titan ocurred. Within
the noise level, no variations were seen.

The composite spectrum is shown in Fig.\,1. It shows several emission molecular signatures attributable to CH$_{4}$, CO, HCN, and H$_{2}$O, molecules that have been previously detected on Titan, but most line transitions are detected
for the first time.
We detected some faint emission features attributable to isotopologues above the noise (3---5\% above the line-to-continuum ratio),
the $^{13}$CO (15--14) and (16--15), and the H$^{13}$CN (19--18) and (20--19)
spectral lines.

\section{Modelling the Titan PACS spectra and comparing them with the observed spectra}
Our simulation package was developed based on the general forward and inversion model
called microwave observation line estimation and retrieval (MOLIERE (v5))\,\citep{u04,ren10, sagawa2010}.
The forward model consists of a line-by-line radiative transfer model that takes into consideration a homogeneous spherically symmetric atmosphere
of Titan (altitudes from 0 to 800\,km) and instrumental
functions (where the Herschel beam profile
and the PACS spectral response are considered, we assumed a uniform brightness temperature on the disk and Titan as a point-source).
CH$_{4}$, CO, HCN, and H$_{2}$O (and their isotopologues) were considered for the line opacity calculation.
Isotopic ratios were initially set by adopting the values derived by \cite{courtin2011} (when available) and terrestrial ratios ones.
For the thermal profile of Titan's atmosphere, we adopted the distribution used in \cite{moreno11},  a combination of
the Huygens Atmospheric Structure Instrument (HASI) profile \citep{ful05} below 140\,km, and Cassini/CIRS stratospheric temperatures
\citep{vina10} above 140\,km.

The \textit{a priori} molecular mixing ratios adopted in this study are those derived by
\cite{niemann2010} and \cite{dekok2007} for CH$_4$ and CO, respectively, and the \textit{a priori} profile is that derived by
\citet{marten2002} for HCN.
Spectroscopic parameters such as the transition frequency and the line intensity were derived from
the HITRAN\,2008 compilation \citep{HITRAN2008}, and line strengths for CH$_{4}$ were taken from \cite{bo10}.
As previously mentioned, we analysed the PACS spectra on the basis of the line-to-continuum ratios.
We used the collision-induced absorption (CIA) coefficients of N$_2$ and CH$_4$ based on the formulation of \citet{bf86}, \citet{bf87}, and \citet{b93}.
A comparison between our forward model and that reported in \citet{moreno2012} under the condition of a pencil beam (a one-dimensional beam, i.e., the beam width is zero)  in the nadir viewing geometry shows that they
agree well within 3\%.
The PACS spectral response modelling ($\lambda/\Delta\lambda$ $\sim$1000--5000, which depends on the wavelength and grating diffraction conditions)
was performed
by following the descriptions by \citet{PACS}, and the related documents on the Herschel Science Center$^1$.
A comparison between observed data and synthetic spectra is shown in Fig.\,1.

The best-fit between the observed and modelled spectra was performed simultaneously by least-squares fitting, that is, minimising $\chi^2$.
As sources of uncertainties we considered measurement and modelling errors.
While the first one was considered as the standard deviation of several ObsIds (when available) and the
oscillations about the smooth spectral continuum, modelling errors were considered as the variance caused by the residual between the
measured and fitted fluxes,
set it with a value of 3\% and 1\% for the blue and red bands, respectively.
We quote our results at 1-$\sigma$ significance level, which is given by a change
$\Delta\chi^2$ = $\chi^2$ - $\chi^2_{min}$ of 1.

\section{Mixing ratios of individual trace constituents}

\noindent\textbf{CH$_4$}: CH$_4$ is the second-most abundant gas in Titan's atmosphere, and its origin is unknown. Possible scenarios
range from a continuous resupply to the atmosphere from a reservoir (from Titan's interior) to a recent
and transitory component of the atmosphere \citep{nixon2012}.
The CH$_4$ distribution has been determined previously, it is uniformly mixed in the entire stratosphere \citep{niemann2005}, and abundances of 1.6 $\pm$ 0.5\%, 1.48 $\pm$ 0.09\% between 76 and 140\,km,  and 1.33 $\pm$ 0.07\% were measured previously
\citep[][]{flasar2005, niemann2010, courtin2011}, respectively.
We used selected CH$_4$ lines, simulated the observed spectra using the forward model (adopting several constant abundance distributions), and determined the best-fit (Fig.\,2-3). We found a best-fit volume-mixing ratio of 1.29 $\pm$ 0.03\%, which is consistent with the SPIRE determination.

  \begin{figure}[t]
  \setcounter{figure}{1}
   \centering
\includegraphics[bb=27 166 571 700,width=0.95\linewidth]{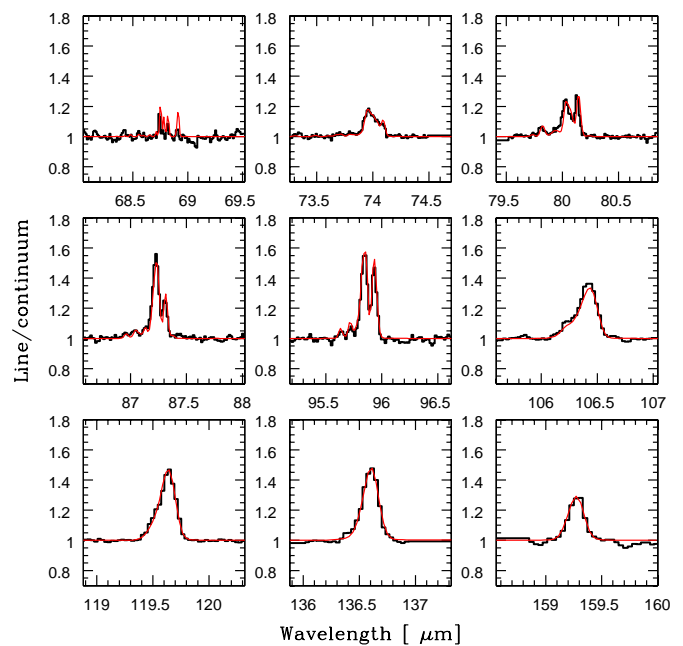}
\caption{Observed and best-fit simulated CH$_4$ lines (black and red, respectively).\label{fig:fig2}}
\end{figure}

  \begin{figure}[h]
  \setcounter{figure}{2}
   \centering
\epsfig{bb = 27 166 571 700, file=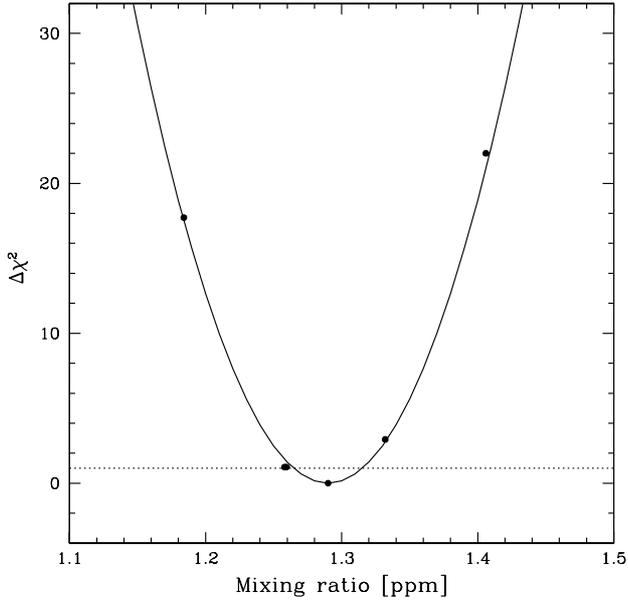,width=0.9\linewidth}
   \caption{Best-fit volume-mixing ratio and the 1-$\sigma$ confidence limits.\label{fig:fig3}}
   \end{figure}

\noindent\textbf{CO}: CO is the fourth-most abundant molecule in Titan's atmosphere. It is not established whether the CO in Titan's atmosphere is primordial (from the interior or surface \citep{wa2004})
or due solely to external sources.
As first shown by \citet{h2008}, the influx of O$^{+}$ from Saturn's magnetosphere into Titan \citep{h06} can lead to the formation of CO through reactions between ground-state O($^3$P) and CH$_3$.
Recent observations suggest that CO in Titan is uniformly mixed
throughout Titan's atmosphere \citep{gurwell2004, baines2006, dekok2007}. Such a profile can be expected if its long chemical lifetime in the oxygen-poor Titan atmosphere is considered
($\sim$10$^{9}$ years \citep{yung1984}). In this study, we retrieved the CO abundance by assuming a
vertically constant profile and simultaneously computed synthetic selected CO lines for several abundances.
We found a best-fit volume-mixing ratio of 50 $\pm$ 2\,ppm (Fig.\,4), consistent with the results from SPIRE/Herschel of 40 $\pm$ 5\,ppm \citep{courtin2011}, Cassini/CIRS \citep{dekok2007} of 47 $\pm$ 8\,ppm, the Atacama Pathfinder EXperiment (APEX)/ Swedish Heterodyne Facility Instrument (SHeFI) of 30$^{+15}_{-8}$\,ppm \citep{ren11}, and the Submillimeter Array (SMA) of 51 $\pm$ 4\,ppm \citep{gurwell2012}.
The PACS CO lines reported here are sensitive to the [60--170]\,km range altitude.

   \begin{figure}[ht]
   \setcounter{figure}{3}
   \centering
   \epsfig{bb = 18 144 592 718,file=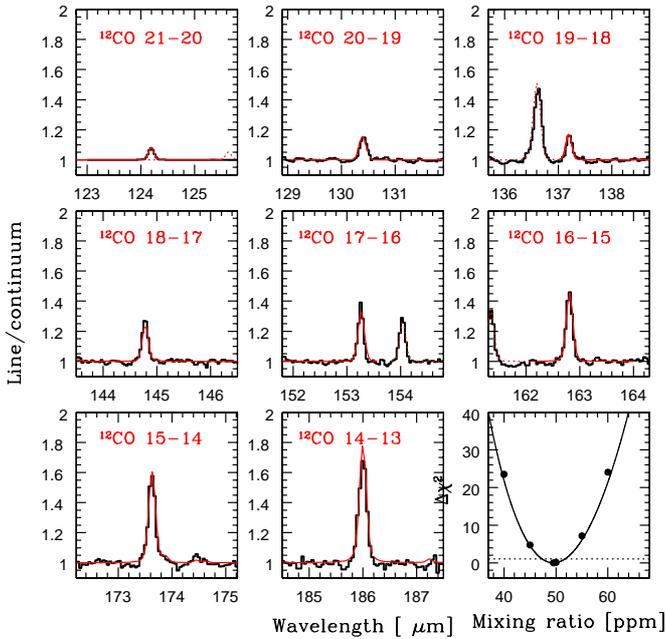,width=1\linewidth,clip=}
   \caption{Observed and best-fit simulated CO lines (black and red, respectively). The dashed lines show the best-fitting simulated lines of other species. Note that the observed CO line at 130.37 $\mu$m is overlapped by the HCN line at 130.24 $\mu$m.
\emph{Bottom-right panel}: best-fit volume-mixing ratio and the 1-$\sigma$ confidence limits.
}
\label{fig:fig4}\end{figure}

\noindent\textbf{HCN}: HCN is generated photochemically in Titan's atmosphere from reactions of hydrocarbon radicals with atomic nitrogen. The latter is produced from EUV or electron impact on N$_2$,
or possibly liberated as a result of cometary impacts
\citep{se11}.
HCN is produced above 300\,km \citep{lara96,wa2004} and removed by
condensation deeper in the atmosphere. This sets up a concentration gradient. A more recent alternative suggests that
HCN is thermodynamically generated via a shock chemistry under lightning discharges in the low atmosphere \citep{kovas2010}.
We analysed the observed spectra by scaling a well-probed reference distribution with a vertically constant factor.
This offers a reliable result under the current data quality
and has also been successfully applied for the Herschel/SPIRE observations \citep{courtin2011},
which also facilitates the inter-comparisons.
We retrieved the scaling factor applied to the adopted distribution of \cite{marten2002}, minimising $\chi^2$, and using a simultaneous combination of five
HCN lines (selected as good-quality lines in terms of the baseline corrections). We obtained
a scaling factor of 1.14 $\pm$ 0.06
as a best-fit to the observed spectrum (Fig.\,5),
consistent with the value of 1.02 $\pm$ 0.13 derived with SPIRE \citep{courtin2011}.
For comparisons the simulated lines following the distribution obtained by Cassini/CIRS \citep{vinatier2007} are also shown.
The CIRS distribution disagrees with the PACS observations at the 1-$\sigma$ level.
We discarded the difference in geometry as the source of the inconsistency because a disk-averaged spectrum gives preferential weight to low latitudes.
Although CIRS data at southern latitudes show no variation during 3.5 years \citep{teanby09}, temporal variability
as the source of discrepancy is not completely ruled out: some scatter in the HCN values in  \cite{teanby09} is compatible with
the differences we notice, and the PACS and CIRS data were not recorded in the same period.
The absolute abundance in the CIRS data seems to be a possible source because by
multiplying the \citet{vinatier2007} distribution by factor of 1.39 a good fit is recovered.
The analysis presented here confirms the results from \cite{marten2002}.
Fig.\,6 shows our retrieved HCN vertical distribution (sensitive to an altitude between 100 and 200\,km)
compared with the distributions obtained with Cassini/CIRS and Herschel/SPIRE.

  \begin{figure}[h]
  \setcounter{figure}{4}
   \centering
\epsfig{file=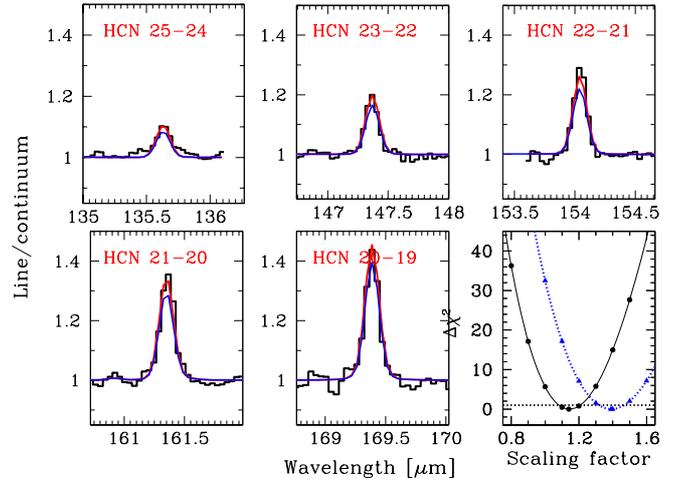,angle=-90, width=1\linewidth}
   \caption{
Observed and best-fit simulated HCN lines (black and red, respectively).
   Blue spectra indicate the simulated spectra using the abundance profile derived from \cite{vinatier2007}\label{fig:fig6}}.
   \end{figure}

  \begin{figure}[h]
  \setcounter{figure}{5}
   \centering
\epsfig{bb = 18 144 592 718, file=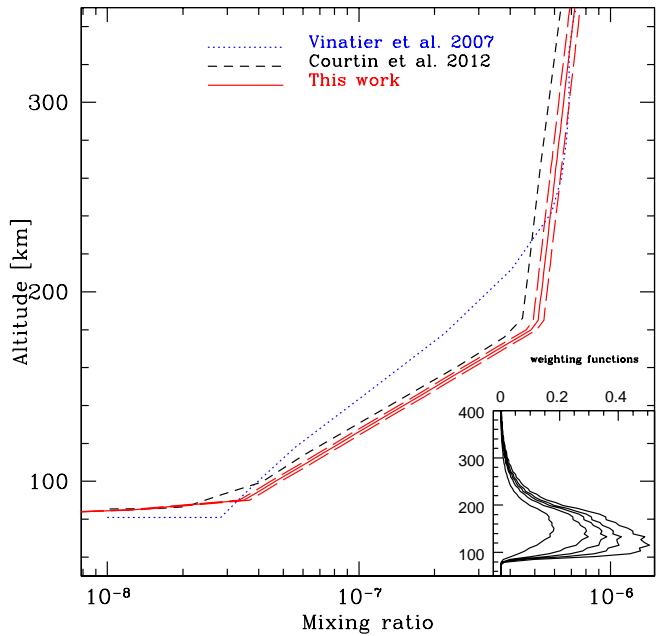,width=1\linewidth}
   \caption{Distribution of HCN (solid line) compared with the profile obtained
   by \cite{vinatier2007} from Cassini/CIRS data at 15$^\circ$S latitude (dotted line) and by
   \cite{courtin2011} from Herschel/SPIRE data (short dashes). Long-dashed lines show the
   1-$\sigma$ limit. The curves in the small box represent the weighting functions for each line. We used the same line order as in Fig.\,5 (from left to right).\label{fig:fig7}}
   \end{figure}

\noindent\textbf{H$_2$O}: The origin of Titan's water remains puzzling. Based on HIFI observations of the H$_2$O Enceladus torus,
\cite{hartogh2011} found that Enceladus is a source for Titan water, but not sufficient given the early estimates of the
influx OH/H$_2$O fluxes into Titan as summarized by
\cite{strobel2010}.
By using combined PACS and HIFI observations of water
lines in Titan, covering several periods in 2010 and 2011 and with a higher S/N,
\citet{moreno2012} determined the H$_2$O profile, and their estimates of the OH/H$_2$O
now offer a possible reconciliation with the input flux into Titan caused by the plume activity of Enceladus. Measurements of water vapor in Titan's
stratosphere have also been recently reported using Cassini/CIRS \citep{cotini2012}. Both Cassini and
Herschel observations indicate that the previous models over-predict the water abundance in Titan's
lower stratosphere, but the two vertical abundance differ. The photochemical model developed by
\cite{k12} yields a H$_2$O profile that includes the O$^+$ flux. This
profile lies between the Herschel \citep{moreno2012} and Cassini \citep{cotini2012}
observations.
We measured the transitions H$_2$O\,(3$_{30}$-2$_{21}$), H$_2$O\,(3$_{31}$-2$_{20}$), and H$_2$O (3$_{21}$-2$_{12}$) at 66.6, 67.1 and 75.4\,$\mu$m, respectively.
The  H$_2$O\,(3$_{31}$-2$_{20}$) one is measured for first time. We simulated the spectra considering the
the semi-empirical S$_a$ profile determined in
\cite{moreno2012}.
 Fig.\,7 shows a comparison between the observed and modelled H$_2$O lines. S$_a$ fits the line at 66.4\,$\mu$m, but underestimates the lines at 67.1 and 75.4 $\mu$m
by a factor of $\sim$0.67. Considering the S/N limitations, we focused on the 66.4\,$\mu$m line, and the S$_a$ distribution is compatible with these PACS lines.
Additional sources of discrepancies are the intensity variations within 10\%, the differences of the radiative
description among the forward models, and errors
due to the incomplete radiative description.

\begin{figure}[h]
\setcounter{figure}{6}
\centering
\epsfig{bb = 18 144 592 718, file=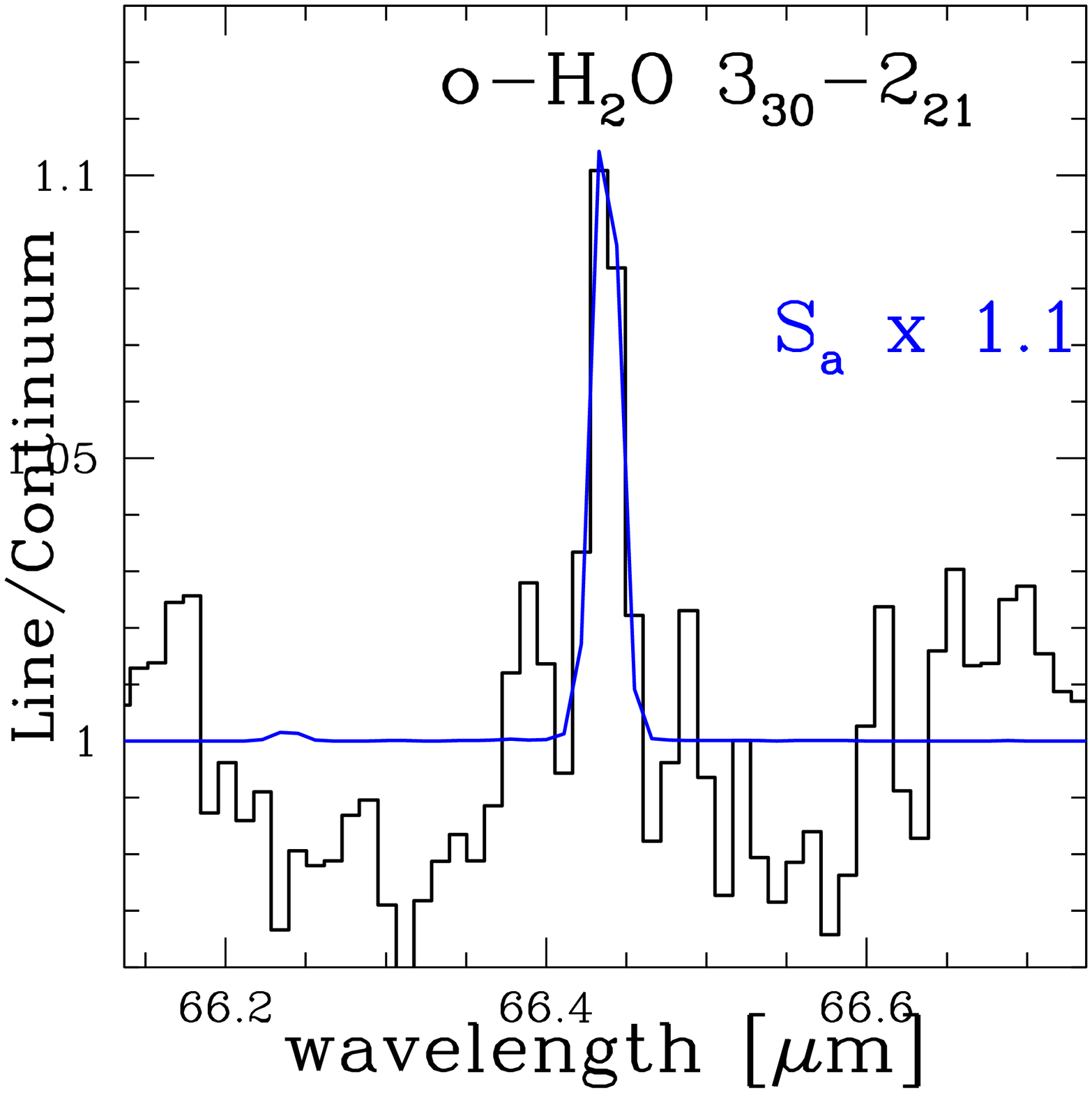,width=0.43\linewidth}
\epsfig{bb = 18 144 592 718, file=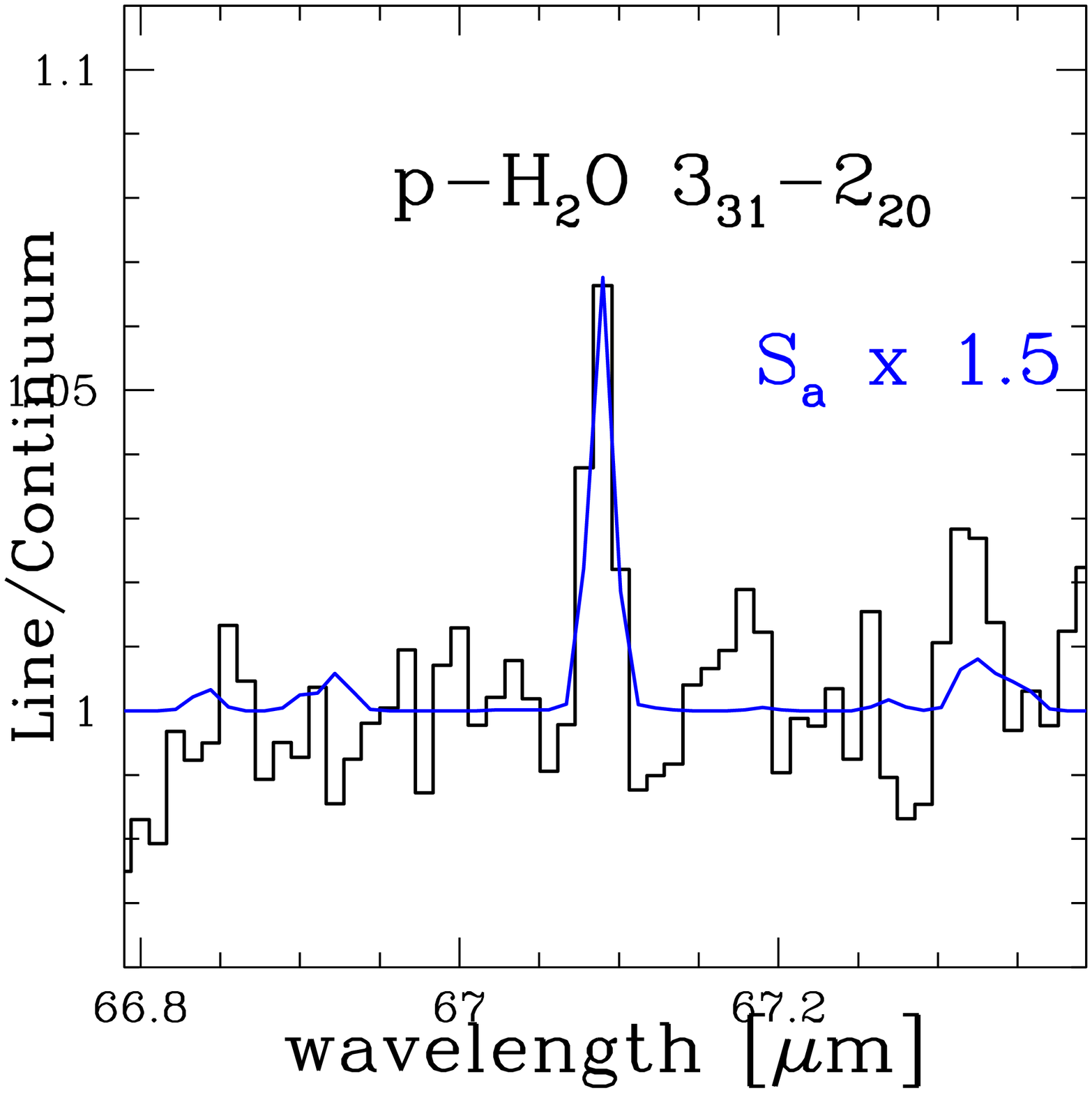,width=0.43\linewidth}
\epsfig{bb = 18 144 592 718, file=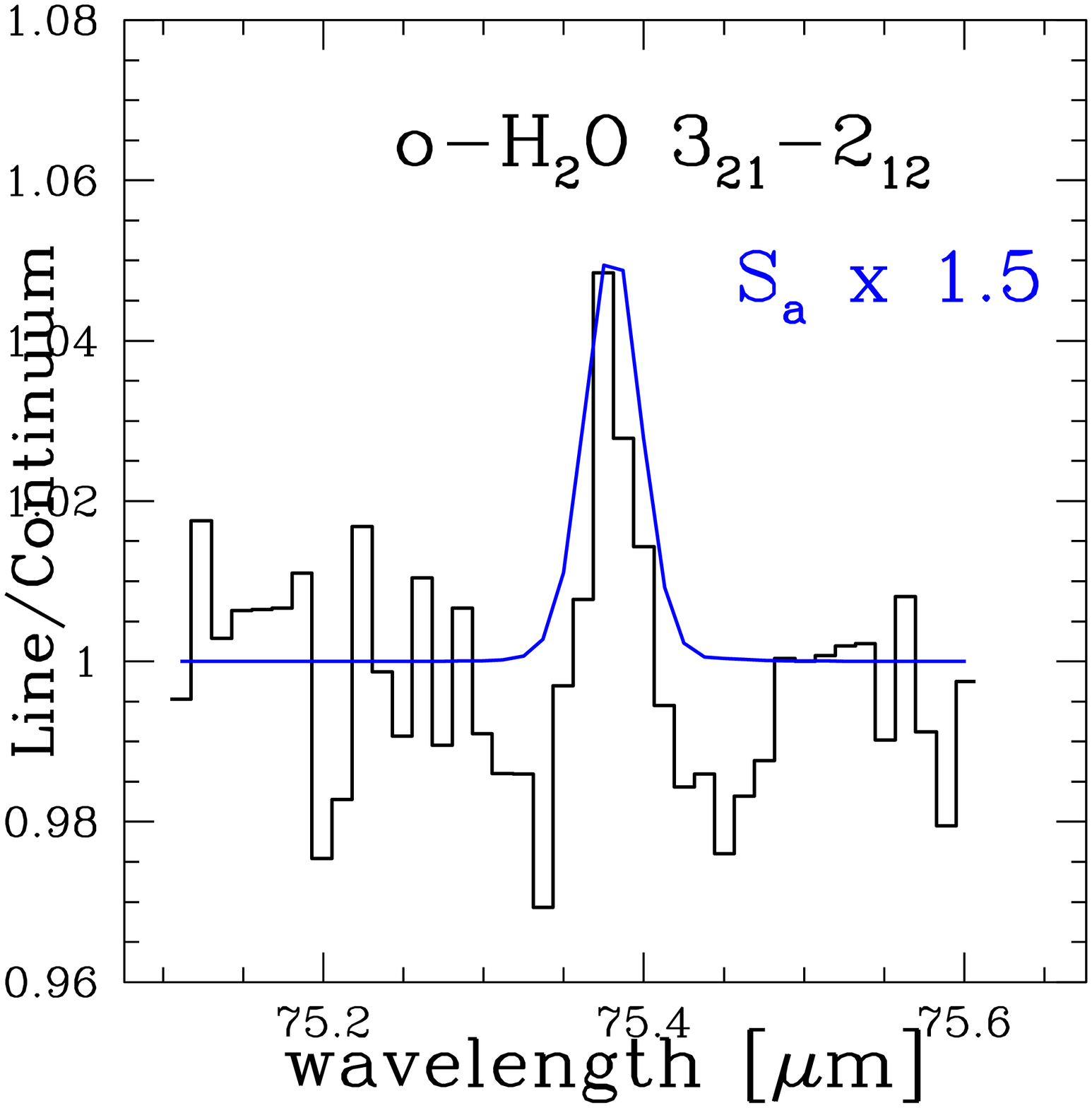,width=0.43\linewidth}
\caption{Comparisons between synthetic spectra computed with the semi-empirical (S$_a$) profile of water \citep{moreno2012} and
an ensemble of three PACS lines at 66.44, 67.09 (ObsId 1342211198) and 75.38\,$\mu$m (ObsId 1342211199).\label{fig:fig8}}
\end{figure}

\noindent\textbf{Isotopic ratios: $^{13}$C/$^{12}$C in CO and HCN}: the detection of the isotopes $^{13}$CO (15--14) and (16--15), and H$^{13}$CN (19--18) and (20--19) is only marginal (Fig.\,1).
Their abundances are found by a least-squares fit, with
errors at the 1-$\sigma$ significance level. The derived isotopic ratio estimates are
$^{12}$C/$^{13}$C of 124 $\pm$ 58, and 66 $\pm$ 35
in CO and HCN, respectively. They are consistent with the terrestrial value of 89.3 given the error bars.
The $^{12}$C/$^{13}$C from this work is consistent with the measurements by SPIRE, CIRS, and SMA \citep{courtin2011, vinatier2007,
gurwell2004, gurwell2008}.

\section{Conclusions}

We reported the first observations obtained with the full range of Herschel/PACS of
Titan. PACS produces a new survey of Titan's spectrum between 51 and 220\,$\mu$m. Signatures attributable to CH$_4$, CO, HCN, and H$_2$O are detected.

We investigated the CH$_4$, CO, and HCN composition of Titan's stratosphere as seen by Herschel/PACS. Nine lines of
CH$_4$ dominate the PACS spectrum and are consistent with an abundance of 1.29 $\pm$ 0.03\%, which is consistent with the value obtained by SPIRE and that
obtained \textit{in situ} by the Huygens/Gas Chromatograph Mass Spectrometer (GCMS).
Eight rotational lines of CO that dominate the PACS spectrum were selected to determine its abundance. Assuming it is well mixed, its abundance was found to be
50 $\pm$ 2\,ppm, consistent with previous studies. Five lines of HCN were selected to
retrieve the HCN vertical distribution. Our HCN mixing-ratio estimations, with a 1.14 scaling factor of the
\cite{marten2002} distribution between 100 and 200\,km,
are consistent with
those with Herschel/SPIRE. The HCN distribution obtained by CIRS does not seem to reproduce the HCN lines observed with PACS well.
 Within the known sources of errors, the H$_2$O distribution proposed from previous Herschel observations produces a satisfactory fit to three H$_2$O lines
detected. The $^{13}$C isotope was detected in CO and HCN.  We determined $^{12}$C/$^{13}$C  in CO and HCN on Titan (sect.\,4).

Limb observations of Titan's submillimetre radiation (e.g. in the 300–-360, 540–-660, and 1080–-1280\,GHz spectral ranges)
are required to detect more rotational lines, measure rotational lines of isotopes, and winds in regions higher than the 200--1200\,km altitude \citep{lellouch2010}.

\begin{acknowledgements}
      We thank the mission planners for the scheduling of these observations,
      L. Rezac for helpful discussions and the anonymous referee for helping to improve the manuscript.
      PACS has been developed by a consortium of institutes led by MPE (Germany) and including UVIE (Austria);
      KUL, CSL, IMEC (Belgium); CEA, OAMP (France); MPIA (Germany); IFSI, OAP/AOT, OAA/CAISMI, LENS, SISSA
      (Italy); IAC (Spain). This development has been supported by the funding agencies BMVIT (Austria),
      ESA-PRODEX (Belgium), CEA/CNES (France), DLR (Germany), ASI (Italy), and CICT/MCT (Spain).
\end{acknowledgements}

\bibliographystyle{aa} 
\bibliography{db1_ref}

\end{document}

%% file: rgb.tex
\definecolor{AliceBlue}{rgb}{0.94,0.97,1.00}
\definecolor{AntiqueWhite1}{rgb}{1.00,0.94,0.86}
\definecolor{AntiqueWhite2}{rgb}{0.93,0.87,0.80}
\definecolor{AntiqueWhite3}{rgb}{0.80,0.75,0.69}
\definecolor{AntiqueWhite4}{rgb}{0.55,0.51,0.47}
\definecolor{AntiqueWhite}{rgb}{0.98,0.92,0.84}
\definecolor{BlanchedAlmond}{rgb}{1.00,0.92,0.80}
\definecolor{BlueViolet}{rgb}{0.54,0.17,0.89}
\definecolor{CadetBlue1}{rgb}{0.60,0.96,1.00}
\definecolor{CadetBlue2}{rgb}{0.56,0.90,0.93}
\definecolor{CadetBlue3}{rgb}{0.48,0.77,0.80}
\definecolor{CadetBlue4}{rgb}{0.33,0.53,0.55}
\definecolor{CadetBlue}{rgb}{0.37,0.62,0.63}
\definecolor{CornflowerBlue}{rgb}{0.39,0.58,0.93}
\definecolor{DarkBlue}{rgb}{0.00,0.00,0.55}
\definecolor{DarkCyan}{rgb}{0.00,0.55,0.55}
\definecolor{DarkGoldenrod1}{rgb}{1.00,0.73,0.06}
\definecolor{DarkGoldenrod2}{rgb}{0.93,0.68,0.05}
\definecolor{DarkGoldenrod3}{rgb}{0.80,0.58,0.05}
\definecolor{DarkGoldenrod4}{rgb}{0.55,0.40,0.03}
\definecolor{DarkGoldenrod}{rgb}{0.72,0.53,0.04}
\definecolor{DarkGray}{rgb}{0.66,0.66,0.66}
\definecolor{DarkGreen}{rgb}{0.00,0.39,0.00}
\definecolor{DarkGrey}{rgb}{0.66,0.66,0.66}
\definecolor{DarkKhaki}{rgb}{0.74,0.72,0.42}
\definecolor{DarkMagenta}{rgb}{0.55,0.00,0.55}
\definecolor{DarkOliveGreen1}{rgb}{0.79,1.00,0.44}
\definecolor{DarkOliveGreen2}{rgb}{0.74,0.93,0.41}
\definecolor{DarkOliveGreen3}{rgb}{0.64,0.80,0.35}
\definecolor{DarkOliveGreen4}{rgb}{0.43,0.55,0.24}
\definecolor{DarkOliveGreen}{rgb}{0.33,0.42,0.18}
\definecolor{DarkOrange1}{rgb}{1.00,0.50,0.00}
\definecolor{DarkOrange2}{rgb}{0.93,0.46,0.00}
\definecolor{DarkOrange3}{rgb}{0.80,0.40,0.00}
\definecolor{DarkOrange4}{rgb}{0.55,0.27,0.00}
\definecolor{DarkOrange}{rgb}{1.00,0.55,0.00}
\definecolor{DarkOrchid1}{rgb}{0.75,0.24,1.00}
\definecolor{DarkOrchid2}{rgb}{0.70,0.23,0.93}
\definecolor{DarkOrchid3}{rgb}{0.60,0.20,0.80}
\definecolor{DarkOrchid4}{rgb}{0.41,0.13,0.55}
\definecolor{DarkOrchid}{rgb}{0.60,0.20,0.80}
\definecolor{DarkRed}{rgb}{0.55,0.00,0.00}
\definecolor{DarkSalmon}{rgb}{0.91,0.59,0.48}
\definecolor{DarkSeaGreen1}{rgb}{0.76,1.00,0.76}
\definecolor{DarkSeaGreen2}{rgb}{0.71,0.93,0.71}
\definecolor{DarkSeaGreen3}{rgb}{0.61,0.80,0.61}
\definecolor{DarkSeaGreen4}{rgb}{0.41,0.55,0.41}
\definecolor{DarkSeaGreen}{rgb}{0.56,0.74,0.56}
\definecolor{DarkSlateBlue}{rgb}{0.28,0.24,0.55}
\definecolor{DarkSlateGray1}{rgb}{0.59,1.00,1.00}
\definecolor{DarkSlateGray2}{rgb}{0.55,0.93,0.93}
\definecolor{DarkSlateGray3}{rgb}{0.47,0.80,0.80}
\definecolor{DarkSlateGray4}{rgb}{0.32,0.55,0.55}
\definecolor{DarkSlateGray}{rgb}{0.18,0.31,0.31}
\definecolor{DarkSlateGrey}{rgb}{0.18,0.31,0.31}
\definecolor{DarkTurquoise}{rgb}{0.00,0.81,0.82}
\definecolor{DarkViolet}{rgb}{0.58,0.00,0.83}
\definecolor{DeepPink1}{rgb}{1.00,0.08,0.58}
\definecolor{DeepPink2}{rgb}{0.93,0.07,0.54}
\definecolor{DeepPink3}{rgb}{0.80,0.06,0.46}
\definecolor{DeepPink4}{rgb}{0.55,0.04,0.31}
\definecolor{DeepPink}{rgb}{1.00,0.08,0.58}
\definecolor{DeepSkyBlue1}{rgb}{0.00,0.75,1.00}
\definecolor{DeepSkyBlue2}{rgb}{0.00,0.70,0.93}
\definecolor{DeepSkyBlue3}{rgb}{0.00,0.60,0.80}
\definecolor{DeepSkyBlue4}{rgb}{0.00,0.41,0.55}
\definecolor{DeepSkyBlue}{rgb}{0.00,0.75,1.00}
\definecolor{DimGray}{rgb}{0.41,0.41,0.41}
\definecolor{DimGrey}{rgb}{0.41,0.41,0.41}
\definecolor{DodgerBlue1}{rgb}{0.12,0.56,1.00}
\definecolor{DodgerBlue2}{rgb}{0.11,0.53,0.93}
\definecolor{DodgerBlue3}{rgb}{0.09,0.45,0.80}
\definecolor{DodgerBlue4}{rgb}{0.06,0.31,0.55}
\definecolor{DodgerBlue}{rgb}{0.12,0.56,1.00}
\definecolor{FloralWhite}{rgb}{1.00,0.98,0.94}
\definecolor{ForestGreen}{rgb}{0.13,0.55,0.13}
\definecolor{GhostWhite}{rgb}{0.97,0.97,1.00}
\definecolor{GreenYellow}{rgb}{0.68,1.00,0.18}
\definecolor{HotPink1}{rgb}{1.00,0.43,0.71}
\definecolor{HotPink2}{rgb}{0.93,0.42,0.65}
\definecolor{HotPink3}{rgb}{0.80,0.38,0.56}
\definecolor{HotPink4}{rgb}{0.55,0.23,0.38}
\definecolor{HotPink}{rgb}{1.00,0.41,0.71}
\definecolor{IndianRed1}{rgb}{1.00,0.42,0.42}
\definecolor{IndianRed2}{rgb}{0.93,0.39,0.39}
\definecolor{IndianRed3}{rgb}{0.80,0.33,0.33}
\definecolor{IndianRed4}{rgb}{0.55,0.23,0.23}
\definecolor{IndianRed}{rgb}{0.80,0.36,0.36}
\definecolor{LavenderBlush1}{rgb}{1.00,0.94,0.96}
\definecolor{LavenderBlush2}{rgb}{0.93,0.88,0.90}
\definecolor{LavenderBlush3}{rgb}{0.80,0.76,0.77}
\definecolor{LavenderBlush4}{rgb}{0.55,0.51,0.53}
\definecolor{LavenderBlush}{rgb}{1.00,0.94,0.96}
\definecolor{LawnGreen}{rgb}{0.49,0.99,0.00}
\definecolor{LemonChiffon1}{rgb}{1.00,0.98,0.80}
\definecolor{LemonChiffon2}{rgb}{0.93,0.91,0.75}
\definecolor{LemonChiffon3}{rgb}{0.80,0.79,0.65}
\definecolor{LemonChiffon4}{rgb}{0.55,0.54,0.44}
\definecolor{LemonChiffon}{rgb}{1.00,0.98,0.80}
\definecolor{LightBlue1}{rgb}{0.75,0.94,1.00}
\definecolor{LightBlue2}{rgb}{0.70,0.87,0.93}
\definecolor{LightBlue3}{rgb}{0.60,0.75,0.80}
\definecolor{LightBlue4}{rgb}{0.41,0.51,0.55}
\definecolor{LightBlue}{rgb}{0.68,0.85,0.90}
\definecolor{LightCoral}{rgb}{0.94,0.50,0.50}
\definecolor{LightCyan1}{rgb}{0.88,1.00,1.00}
\definecolor{LightCyan2}{rgb}{0.82,0.93,0.93}
\definecolor{LightCyan3}{rgb}{0.71,0.80,0.80}
\definecolor{LightCyan4}{rgb}{0.48,0.55,0.55}
\definecolor{LightCyan}{rgb}{0.88,1.00,1.00}
\definecolor{LightGoldenrod1}{rgb}{1.00,0.93,0.55}
\definecolor{LightGoldenrod2}{rgb}{0.93,0.86,0.51}
\definecolor{LightGoldenrod3}{rgb}{0.80,0.75,0.44}
\definecolor{LightGoldenrod4}{rgb}{0.55,0.51,0.30}
\definecolor{LightGoldenrodYellow}{rgb}{0.98,0.98,0.82}
\definecolor{LightGoldenrod}{rgb}{0.93,0.87,0.51}
\definecolor{LightGray}{rgb}{0.83,0.83,0.83}
\definecolor{LightGreen}{rgb}{0.56,0.93,0.56}
\definecolor{LightGrey}{rgb}{0.83,0.83,0.83}
\definecolor{LightPink1}{rgb}{1.00,0.68,0.73}
\definecolor{LightPink2}{rgb}{0.93,0.64,0.68}
\definecolor{LightPink3}{rgb}{0.80,0.55,0.58}
\definecolor{LightPink4}{rgb}{0.55,0.37,0.40}
\definecolor{LightPink}{rgb}{1.00,0.71,0.76}
\definecolor{LightSalmon1}{rgb}{1.00,0.63,0.48}
\definecolor{LightSalmon2}{rgb}{0.93,0.58,0.45}
\definecolor{LightSalmon3}{rgb}{0.80,0.51,0.38}
\definecolor{LightSalmon4}{rgb}{0.55,0.34,0.26}
\definecolor{LightSalmon}{rgb}{1.00,0.63,0.48}
\definecolor{LightSeaGreen}{rgb}{0.13,0.70,0.67}
\definecolor{LightSkyBlue1}{rgb}{0.69,0.89,1.00}
\definecolor{LightSkyBlue2}{rgb}{0.64,0.83,0.93}
\definecolor{LightSkyBlue3}{rgb}{0.55,0.71,0.80}
\definecolor{LightSkyBlue4}{rgb}{0.38,0.48,0.55}
\definecolor{LightSkyBlue}{rgb}{0.53,0.81,0.98}
\definecolor{LightSlateBlue}{rgb}{0.52,0.44,1.00}
\definecolor{LightSlateGray}{rgb}{0.47,0.53,0.60}
\definecolor{LightSlateGrey}{rgb}{0.47,0.53,0.60}
\definecolor{LightSteelBlue1}{rgb}{0.79,0.88,1.00}
\definecolor{LightSteelBlue2}{rgb}{0.74,0.82,0.93}
\definecolor{LightSteelBlue3}{rgb}{0.64,0.71,0.80}
\definecolor{LightSteelBlue4}{rgb}{0.43,0.48,0.55}
\definecolor{LightSteelBlue}{rgb}{0.69,0.77,0.87}
\definecolor{LightYellow1}{rgb}{1.00,1.00,0.88}
\definecolor{LightYellow2}{rgb}{0.93,0.93,0.82}
\definecolor{LightYellow3}{rgb}{0.80,0.80,0.71}
\definecolor{LightYellow4}{rgb}{0.55,0.55,0.48}
\definecolor{LightYellow}{rgb}{1.00,1.00,0.88}
\definecolor{LimeGreen}{rgb}{0.20,0.80,0.20}
\definecolor{MediumAquamarine}{rgb}{0.40,0.80,0.67}
\definecolor{MediumBlue}{rgb}{0.00,0.00,0.80}
\definecolor{MediumOrchid1}{rgb}{0.88,0.40,1.00}
\definecolor{MediumOrchid2}{rgb}{0.82,0.37,0.93}
\definecolor{MediumOrchid3}{rgb}{0.71,0.32,0.80}
\definecolor{MediumOrchid4}{rgb}{0.48,0.22,0.55}
\definecolor{MediumOrchid}{rgb}{0.73,0.33,0.83}
\definecolor{MediumPurple1}{rgb}{0.67,0.51,1.00}
\definecolor{MediumPurple2}{rgb}{0.62,0.47,0.93}
\definecolor{MediumPurple3}{rgb}{0.54,0.41,0.80}
\definecolor{MediumPurple4}{rgb}{0.36,0.28,0.55}
\definecolor{MediumPurple}{rgb}{0.58,0.44,0.86}
\definecolor{MediumSeaGreen}{rgb}{0.24,0.70,0.44}
\definecolor{MediumSlateBlue}{rgb}{0.48,0.41,0.93}
\definecolor{MediumSpringGreen}{rgb}{0.00,0.98,0.60}
\definecolor{MediumTurquoise}{rgb}{0.28,0.82,0.80}
\definecolor{MediumVioletRed}{rgb}{0.78,0.08,0.52}
\definecolor{MidnightBlue}{rgb}{0.10,0.10,0.44}
\definecolor{MintCream}{rgb}{0.96,1.00,0.98}
\definecolor{MistyRose1}{rgb}{1.00,0.89,0.88}
\definecolor{MistyRose2}{rgb}{0.93,0.84,0.82}
\definecolor{MistyRose3}{rgb}{0.80,0.72,0.71}
\definecolor{MistyRose4}{rgb}{0.55,0.49,0.48}
\definecolor{MistyRose}{rgb}{1.00,0.89,0.88}
\definecolor{NavajoWhite1}{rgb}{1.00,0.87,0.68}
\definecolor{NavajoWhite2}{rgb}{0.93,0.81,0.63}
\definecolor{NavajoWhite3}{rgb}{0.80,0.70,0.55}
\definecolor{NavajoWhite4}{rgb}{0.55,0.47,0.37}
\definecolor{NavajoWhite}{rgb}{1.00,0.87,0.68}
\definecolor{NavyBlue}{rgb}{0.00,0.00,0.50}
\definecolor{OldLace}{rgb}{0.99,0.96,0.90}
\definecolor{OliveDrab1}{rgb}{0.75,1.00,0.24}
\definecolor{OliveDrab2}{rgb}{0.70,0.93,0.23}
\definecolor{OliveDrab3}{rgb}{0.60,0.80,0.20}
\definecolor{OliveDrab4}{rgb}{0.41,0.55,0.13}
\definecolor{OliveDrab}{rgb}{0.42,0.56,0.14}
\definecolor{OrangeRed1}{rgb}{1.00,0.27,0.00}
\definecolor{OrangeRed2}{rgb}{0.93,0.25,0.00}
\definecolor{OrangeRed3}{rgb}{0.80,0.22,0.00}
\definecolor{OrangeRed4}{rgb}{0.55,0.15,0.00}
\definecolor{OrangeRed}{rgb}{1.00,0.27,0.00}
\definecolor{PaleGoldenrod}{rgb}{0.93,0.91,0.67}
\definecolor{PaleGreen1}{rgb}{0.60,1.00,0.60}
\definecolor{PaleGreen2}{rgb}{0.56,0.93,0.56}
\definecolor{PaleGreen3}{rgb}{0.49,0.80,0.49}
\definecolor{PaleGreen4}{rgb}{0.33,0.55,0.33}
\definecolor{PaleGreen}{rgb}{0.60,0.98,0.60}
\definecolor{PaleTurquoise1}{rgb}{0.73,1.00,1.00}
\definecolor{PaleTurquoise2}{rgb}{0.68,0.93,0.93}
\definecolor{PaleTurquoise3}{rgb}{0.59,0.80,0.80}
\definecolor{PaleTurquoise4}{rgb}{0.40,0.55,0.55}
\definecolor{PaleTurquoise}{rgb}{0.69,0.93,0.93}
\definecolor{PaleVioletRed1}{rgb}{1.00,0.51,0.67}
\definecolor{PaleVioletRed2}{rgb}{0.93,0.47,0.62}
\definecolor{PaleVioletRed3}{rgb}{0.80,0.41,0.54}
\definecolor{PaleVioletRed4}{rgb}{0.55,0.28,0.36}
\definecolor{PaleVioletRed}{rgb}{0.86,0.44,0.58}
\definecolor{PapayaWhip}{rgb}{1.00,0.94,0.84}
\definecolor{PeachPuff1}{rgb}{1.00,0.85,0.73}
\definecolor{PeachPuff2}{rgb}{0.93,0.80,0.68}
\definecolor{PeachPuff3}{rgb}{0.80,0.69,0.58}
\definecolor{PeachPuff4}{rgb}{0.55,0.47,0.40}
\definecolor{PeachPuff}{rgb}{1.00,0.85,0.73}
\definecolor{PowderBlue}{rgb}{0.69,0.88,0.90}
\definecolor{RosyBrown1}{rgb}{1.00,0.76,0.76}
\definecolor{RosyBrown2}{rgb}{0.93,0.71,0.71}
\definecolor{RosyBrown3}{rgb}{0.80,0.61,0.61}
\definecolor{RosyBrown4}{rgb}{0.55,0.41,0.41}
\definecolor{RosyBrown}{rgb}{0.74,0.56,0.56}
\definecolor{RoyalBlue1}{rgb}{0.28,0.46,1.00}
\definecolor{RoyalBlue2}{rgb}{0.26,0.43,0.93}
\definecolor{RoyalBlue3}{rgb}{0.23,0.37,0.80}
\definecolor{RoyalBlue4}{rgb}{0.15,0.25,0.55}
\definecolor{RoyalBlue}{rgb}{0.25,0.41,0.88}
\definecolor{SaddleBrown}{rgb}{0.55,0.27,0.07}
\definecolor{SandyBrown}{rgb}{0.96,0.64,0.38}
\definecolor{SeaGreen1}{rgb}{0.33,1.00,0.62}
\definecolor{SeaGreen2}{rgb}{0.31,0.93,0.58}
\definecolor{SeaGreen3}{rgb}{0.26,0.80,0.50}
\definecolor{SeaGreen4}{rgb}{0.18,0.55,0.34}
\definecolor{SeaGreen}{rgb}{0.18,0.55,0.34}
\definecolor{SkyBlue1}{rgb}{0.53,0.81,1.00}
\definecolor{SkyBlue2}{rgb}{0.49,0.75,0.93}
\definecolor{SkyBlue3}{rgb}{0.42,0.65,0.80}
\definecolor{SkyBlue4}{rgb}{0.29,0.44,0.55}
\definecolor{SkyBlue}{rgb}{0.53,0.81,0.92}
\definecolor{SlateBlue1}{rgb}{0.51,0.44,1.00}
\definecolor{SlateBlue2}{rgb}{0.48,0.40,0.93}
\definecolor{SlateBlue3}{rgb}{0.41,0.35,0.80}
\definecolor{SlateBlue4}{rgb}{0.28,0.24,0.55}
\definecolor{SlateBlue}{rgb}{0.42,0.35,0.80}
\definecolor{SlateGray1}{rgb}{0.78,0.89,1.00}
\definecolor{SlateGray2}{rgb}{0.73,0.83,0.93}
\definecolor{SlateGray3}{rgb}{0.62,0.71,0.80}
\definecolor{SlateGray4}{rgb}{0.42,0.48,0.55}
\definecolor{SlateGray}{rgb}{0.44,0.50,0.56}
\definecolor{SlateGrey}{rgb}{0.44,0.50,0.56}
\definecolor{SpringGreen1}{rgb}{0.00,1.00,0.50}
\definecolor{SpringGreen2}{rgb}{0.00,0.93,0.46}
\definecolor{SpringGreen3}{rgb}{0.00,0.80,0.40}
\definecolor{SpringGreen4}{rgb}{0.00,0.55,0.27}
\definecolor{SpringGreen}{rgb}{0.00,1.00,0.50}
\definecolor{SteelBlue1}{rgb}{0.39,0.72,1.00}
\definecolor{SteelBlue2}{rgb}{0.36,0.67,0.93}
\definecolor{SteelBlue3}{rgb}{0.31,0.58,0.80}
\definecolor{SteelBlue4}{rgb}{0.21,0.39,0.55}
\definecolor{SteelBlue}{rgb}{0.27,0.51,0.71}
\definecolor{VioletRed1}{rgb}{1.00,0.24,0.59}
\definecolor{VioletRed2}{rgb}{0.93,0.23,0.55}
\definecolor{VioletRed3}{rgb}{0.80,0.20,0.47}
\definecolor{VioletRed4}{rgb}{0.55,0.13,0.32}
\definecolor{VioletRed}{rgb}{0.82,0.13,0.56}
\definecolor{WhiteSmoke}{rgb}{0.96,0.96,0.96}
\definecolor{YellowGreen}{rgb}{0.60,0.80,0.20}
\definecolor{aliceblue}{rgb}{0.94,0.97,1.00}
\definecolor{antiquewhite}{rgb}{0.98,0.92,0.84}
\definecolor{aquamarine1}{rgb}{0.50,1.00,0.83}
\definecolor{aquamarine2}{rgb}{0.46,0.93,0.78}
\definecolor{aquamarine3}{rgb}{0.40,0.80,0.67}
\definecolor{aquamarine4}{rgb}{0.27,0.55,0.45}
\definecolor{aquamarine}{rgb}{0.50,1.00,0.83}
\definecolor{azure1}{rgb}{0.94,1.00,1.00}
\definecolor{azure2}{rgb}{0.88,0.93,0.93}
\definecolor{azure3}{rgb}{0.76,0.80,0.80}
\definecolor{azure4}{rgb}{0.51,0.55,0.55}
\definecolor{azure}{rgb}{0.94,1.00,1.00}
\definecolor{beige}{rgb}{0.96,0.96,0.86}
\definecolor{bisque1}{rgb}{1.00,0.89,0.77}
\definecolor{bisque2}{rgb}{0.93,0.84,0.72}
\definecolor{bisque3}{rgb}{0.80,0.72,0.62}
\definecolor{bisque4}{rgb}{0.55,0.49,0.42}
\definecolor{bisque}{rgb}{1.00,0.89,0.77}
\definecolor{black}{rgb}{0.00,0.00,0.00}
\definecolor{blanchedalmond}{rgb}{1.00,0.92,0.80}
\definecolor{blue1}{rgb}{0.00,0.00,1.00}
\definecolor{blue2}{rgb}{0.00,0.00,0.93}
\definecolor{blue3}{rgb}{0.00,0.00,0.80}
\definecolor{blue4}{rgb}{0.00,0.00,0.55}
\definecolor{blueviolet}{rgb}{0.54,0.17,0.89}
\definecolor{blue}{rgb}{0.00,0.00,1.00}
\definecolor{brown1}{rgb}{1.00,0.25,0.25}
\definecolor{brown2}{rgb}{0.93,0.23,0.23}
\definecolor{brown3}{rgb}{0.80,0.20,0.20}
\definecolor{brown4}{rgb}{0.55,0.14,0.14}
\definecolor{brown}{rgb}{0.65,0.16,0.16}
\definecolor{burlywood1}{rgb}{1.00,0.83,0.61}
\definecolor{burlywood2}{rgb}{0.93,0.77,0.57}
\definecolor{burlywood3}{rgb}{0.80,0.67,0.49}
\definecolor{burlywood4}{rgb}{0.55,0.45,0.33}
\definecolor{burlywood}{rgb}{0.87,0.72,0.53}
\definecolor{cadetblue}{rgb}{0.37,0.62,0.63}
\definecolor{chartreuse1}{rgb}{0.50,1.00,0.00}
\definecolor{chartreuse2}{rgb}{0.46,0.93,0.00}
\definecolor{chartreuse3}{rgb}{0.40,0.80,0.00}
\definecolor{chartreuse4}{rgb}{0.27,0.55,0.00}
\definecolor{chartreuse}{rgb}{0.50,1.00,0.00}
\definecolor{chocolate1}{rgb}{1.00,0.50,0.14}
\definecolor{chocolate2}{rgb}{0.93,0.46,0.13}
\definecolor{chocolate3}{rgb}{0.80,0.40,0.11}
\definecolor{chocolate4}{rgb}{0.55,0.27,0.07}
\definecolor{chocolate}{rgb}{0.82,0.41,0.12}
\definecolor{coral1}{rgb}{1.00,0.45,0.34}
\definecolor{coral2}{rgb}{0.93,0.42,0.31}
\definecolor{coral3}{rgb}{0.80,0.36,0.27}
\definecolor{coral4}{rgb}{0.55,0.24,0.18}
\definecolor{coral}{rgb}{1.00,0.50,0.31}
\definecolor{cornflowerblue}{rgb}{0.39,0.58,0.93}
\definecolor{cornsilk1}{rgb}{1.00,0.97,0.86}
\definecolor{cornsilk2}{rgb}{0.93,0.91,0.80}
\definecolor{cornsilk3}{rgb}{0.80,0.78,0.69}
\definecolor{cornsilk4}{rgb}{0.55,0.53,0.47}
\definecolor{cornsilk}{rgb}{1.00,0.97,0.86}
\definecolor{cyan1}{rgb}{0.00,1.00,1.00}
\definecolor{cyan2}{rgb}{0.00,0.93,0.93}
\definecolor{cyan3}{rgb}{0.00,0.80,0.80}
\definecolor{cyan4}{rgb}{0.00,0.55,0.55}
\definecolor{cyan}{rgb}{0.00,1.00,1.00}
\definecolor{darkblue}{rgb}{0.00,0.00,0.55}
\definecolor{darkcyan}{rgb}{0.00,0.55,0.55}
\definecolor{darkgoldenrod}{rgb}{0.72,0.53,0.04}
\definecolor{darkgray}{rgb}{0.66,0.66,0.66}
\definecolor{darkgreen}{rgb}{0.00,0.39,0.00}
\definecolor{darkgrey}{rgb}{0.66,0.66,0.66}
\definecolor{darkkhaki}{rgb}{0.74,0.72,0.42}
\definecolor{darkmagenta}{rgb}{0.55,0.00,0.55}
\definecolor{darkolive}{rgb}{0.33,0.42,0.18}
\definecolor{darkorange}{rgb}{1.00,0.55,0.00}
\definecolor{darkorchid}{rgb}{0.60,0.20,0.80}
\definecolor{darkred}{rgb}{0.55,0.00,0.00}
\definecolor{darksalmon}{rgb}{0.91,0.59,0.48}
\definecolor{darksea}{rgb}{0.56,0.74,0.56}
\definecolor{darkslate}{rgb}{0.18,0.31,0.31}
\definecolor{darkslate}{rgb}{0.18,0.31,0.31}
\definecolor{darkslate}{rgb}{0.28,0.24,0.55}
\definecolor{darkturquoise}{rgb}{0.00,0.81,0.82}
\definecolor{darkviolet}{rgb}{0.58,0.00,0.83}
\definecolor{deeppink}{rgb}{1.00,0.08,0.58}
\definecolor{deepsky}{rgb}{0.00,0.75,1.00}
\definecolor{dimgray}{rgb}{0.41,0.41,0.41}
\definecolor{dimgrey}{rgb}{0.41,0.41,0.41}
\definecolor{dodgerblue}{rgb}{0.12,0.56,1.00}
\definecolor{firebrick1}{rgb}{1.00,0.19,0.19}
\definecolor{firebrick2}{rgb}{0.93,0.17,0.17}
\definecolor{firebrick3}{rgb}{0.80,0.15,0.15}
\definecolor{firebrick4}{rgb}{0.55,0.10,0.10}
\definecolor{firebrick}{rgb}{0.70,0.13,0.13}
\definecolor{floralwhite}{rgb}{1.00,0.98,0.94}
\definecolor{forestgreen}{rgb}{0.13,0.55,0.13}
\definecolor{gainsboro}{rgb}{0.86,0.86,0.86}
\definecolor{ghostwhite}{rgb}{0.97,0.97,1.00}
\definecolor{gold1}{rgb}{1.00,0.84,0.00}
\definecolor{gold2}{rgb}{0.93,0.79,0.00}
\definecolor{gold3}{rgb}{0.80,0.68,0.00}
\definecolor{gold4}{rgb}{0.55,0.46,0.00}
\definecolor{goldenrod1}{rgb}{1.00,0.76,0.15}
\definecolor{goldenrod2}{rgb}{0.93,0.71,0.13}
\definecolor{goldenrod3}{rgb}{0.80,0.61,0.11}
\definecolor{goldenrod4}{rgb}{0.55,0.41,0.08}
\definecolor{goldenrod}{rgb}{0.85,0.65,0.13}
\definecolor{gold}{rgb}{1.00,0.84,0.00}
\definecolor{gray0}{rgb}{0.00,0.00,0.00}
\definecolor{gray100}{rgb}{1.00,1.00,1.00}
\definecolor{gray10}{rgb}{0.10,0.10,0.10}
\definecolor{gray11}{rgb}{0.11,0.11,0.11}
\definecolor{gray12}{rgb}{0.12,0.12,0.12}
\definecolor{gray13}{rgb}{0.13,0.13,0.13}
\definecolor{gray14}{rgb}{0.14,0.14,0.14}
\definecolor{gray15}{rgb}{0.15,0.15,0.15}
\definecolor{gray16}{rgb}{0.16,0.16,0.16}
\definecolor{gray17}{rgb}{0.17,0.17,0.17}
\definecolor{gray18}{rgb}{0.18,0.18,0.18}
\definecolor{gray19}{rgb}{0.19,0.19,0.19}
\definecolor{gray1}{rgb}{0.01,0.01,0.01}
\definecolor{gray20}{rgb}{0.20,0.20,0.20}
\definecolor{gray21}{rgb}{0.21,0.21,0.21}
\definecolor{gray22}{rgb}{0.22,0.22,0.22}
\definecolor{gray23}{rgb}{0.23,0.23,0.23}
\definecolor{gray24}{rgb}{0.24,0.24,0.24}
\definecolor{gray25}{rgb}{0.25,0.25,0.25}
\definecolor{gray26}{rgb}{0.26,0.26,0.26}
\definecolor{gray27}{rgb}{0.27,0.27,0.27}
\definecolor{gray28}{rgb}{0.28,0.28,0.28}
\definecolor{gray29}{rgb}{0.29,0.29,0.29}
\definecolor{gray2}{rgb}{0.02,0.02,0.02}
\definecolor{gray30}{rgb}{0.30,0.30,0.30}
\definecolor{gray31}{rgb}{0.31,0.31,0.31}
\definecolor{gray32}{rgb}{0.32,0.32,0.32}
\definecolor{gray33}{rgb}{0.33,0.33,0.33}
\definecolor{gray34}{rgb}{0.34,0.34,0.34}
\definecolor{gray35}{rgb}{0.35,0.35,0.35}
\definecolor{gray36}{rgb}{0.36,0.36,0.36}
\definecolor{gray37}{rgb}{0.37,0.37,0.37}
\definecolor{gray38}{rgb}{0.38,0.38,0.38}
\definecolor{gray39}{rgb}{0.39,0.39,0.39}
\definecolor{gray3}{rgb}{0.03,0.03,0.03}
\definecolor{gray40}{rgb}{0.40,0.40,0.40}
\definecolor{gray41}{rgb}{0.41,0.41,0.41}
\definecolor{gray42}{rgb}{0.42,0.42,0.42}
\definecolor{gray43}{rgb}{0.43,0.43,0.43}
\definecolor{gray44}{rgb}{0.44,0.44,0.44}
\definecolor{gray45}{rgb}{0.45,0.45,0.45}
\definecolor{gray46}{rgb}{0.46,0.46,0.46}
\definecolor{gray47}{rgb}{0.47,0.47,0.47}
\definecolor{gray48}{rgb}{0.48,0.48,0.48}
\definecolor{gray49}{rgb}{0.49,0.49,0.49}
\definecolor{gray4}{rgb}{0.04,0.04,0.04}
\definecolor{gray50}{rgb}{0.50,0.50,0.50}
\definecolor{gray51}{rgb}{0.51,0.51,0.51}
\definecolor{gray52}{rgb}{0.52,0.52,0.52}
\definecolor{gray53}{rgb}{0.53,0.53,0.53}
\definecolor{gray54}{rgb}{0.54,0.54,0.54}
\definecolor{gray55}{rgb}{0.55,0.55,0.55}
\definecolor{gray56}{rgb}{0.56,0.56,0.56}
\definecolor{gray57}{rgb}{0.57,0.57,0.57}
\definecolor{gray58}{rgb}{0.58,0.58,0.58}
\definecolor{gray59}{rgb}{0.59,0.59,0.59}
\definecolor{gray5}{rgb}{0.05,0.05,0.05}
\definecolor{gray60}{rgb}{0.60,0.60,0.60}
\definecolor{gray61}{rgb}{0.61,0.61,0.61}
\definecolor{gray62}{rgb}{0.62,0.62,0.62}
\definecolor{gray63}{rgb}{0.63,0.63,0.63}
\definecolor{gray64}{rgb}{0.64,0.64,0.64}
\definecolor{gray65}{rgb}{0.65,0.65,0.65}
\definecolor{gray66}{rgb}{0.66,0.66,0.66}
\definecolor{gray67}{rgb}{0.67,0.67,0.67}
\definecolor{gray68}{rgb}{0.68,0.68,0.68}
\definecolor{gray69}{rgb}{0.69,0.69,0.69}
\definecolor{gray6}{rgb}{0.06,0.06,0.06}
\definecolor{gray70}{rgb}{0.70,0.70,0.70}
\definecolor{gray71}{rgb}{0.71,0.71,0.71}
\definecolor{gray72}{rgb}{0.72,0.72,0.72}
\definecolor{gray73}{rgb}{0.73,0.73,0.73}
\definecolor{gray74}{rgb}{0.74,0.74,0.74}
\definecolor{gray75}{rgb}{0.75,0.75,0.75}
\definecolor{gray76}{rgb}{0.76,0.76,0.76}
\definecolor{gray77}{rgb}{0.77,0.77,0.77}
\definecolor{gray78}{rgb}{0.78,0.78,0.78}
\definecolor{gray79}{rgb}{0.79,0.79,0.79}
\definecolor{gray7}{rgb}{0.07,0.07,0.07}
\definecolor{gray80}{rgb}{0.80,0.80,0.80}
\definecolor{gray81}{rgb}{0.81,0.81,0.81}
\definecolor{gray82}{rgb}{0.82,0.82,0.82}
\definecolor{gray83}{rgb}{0.83,0.83,0.83}
\definecolor{gray84}{rgb}{0.84,0.84,0.84}
\definecolor{gray85}{rgb}{0.85,0.85,0.85}
\definecolor{gray86}{rgb}{0.86,0.86,0.86}
\definecolor{gray87}{rgb}{0.87,0.87,0.87}
\definecolor{gray88}{rgb}{0.88,0.88,0.88}
\definecolor{gray89}{rgb}{0.89,0.89,0.89}
\definecolor{gray8}{rgb}{0.08,0.08,0.08}
\definecolor{gray90}{rgb}{0.90,0.90,0.90}
\definecolor{gray91}{rgb}{0.91,0.91,0.91}
\definecolor{gray92}{rgb}{0.92,0.92,0.92}
\definecolor{gray93}{rgb}{0.93,0.93,0.93}
\definecolor{gray94}{rgb}{0.94,0.94,0.94}
\definecolor{gray95}{rgb}{0.95,0.95,0.95}
\definecolor{gray96}{rgb}{0.96,0.96,0.96}
\definecolor{gray97}{rgb}{0.97,0.97,0.97}
\definecolor{gray98}{rgb}{0.98,0.98,0.98}
\definecolor{gray99}{rgb}{0.99,0.99,0.99}
\definecolor{gray9}{rgb}{0.09,0.09,0.09}
\definecolor{gray}{rgb}{0.75,0.75,0.75}
\definecolor{green1}{rgb}{0.00,1.00,0.00}
\definecolor{green2}{rgb}{0.00,0.93,0.00}
\definecolor{green3}{rgb}{0.00,0.80,0.00}
\definecolor{green4}{rgb}{0.00,0.55,0.00}
\definecolor{greenyellow}{rgb}{0.68,1.00,0.18}
\definecolor{green}{rgb}{0.00,1.00,0.00}
\definecolor{grey0}{rgb}{0.00,0.00,0.00}
\definecolor{grey100}{rgb}{1.00,1.00,1.00}
\definecolor{grey10}{rgb}{0.10,0.10,0.10}
\definecolor{grey11}{rgb}{0.11,0.11,0.11}
\definecolor{grey12}{rgb}{0.12,0.12,0.12}
\definecolor{grey13}{rgb}{0.13,0.13,0.13}
\definecolor{grey14}{rgb}{0.14,0.14,0.14}
\definecolor{grey15}{rgb}{0.15,0.15,0.15}
\definecolor{grey16}{rgb}{0.16,0.16,0.16}
\definecolor{grey17}{rgb}{0.17,0.17,0.17}
\definecolor{grey18}{rgb}{0.18,0.18,0.18}
\definecolor{grey19}{rgb}{0.19,0.19,0.19}
\definecolor{grey1}{rgb}{0.01,0.01,0.01}
\definecolor{grey20}{rgb}{0.20,0.20,0.20}
\definecolor{grey21}{rgb}{0.21,0.21,0.21}
\definecolor{grey22}{rgb}{0.22,0.22,0.22}
\definecolor{grey23}{rgb}{0.23,0.23,0.23}
\definecolor{grey24}{rgb}{0.24,0.24,0.24}
\definecolor{grey25}{rgb}{0.25,0.25,0.25}
\definecolor{grey26}{rgb}{0.26,0.26,0.26}
\definecolor{grey27}{rgb}{0.27,0.27,0.27}
\definecolor{grey28}{rgb}{0.28,0.28,0.28}
\definecolor{grey29}{rgb}{0.29,0.29,0.29}
\definecolor{grey2}{rgb}{0.02,0.02,0.02}
\definecolor{grey30}{rgb}{0.30,0.30,0.30}
\definecolor{grey31}{rgb}{0.31,0.31,0.31}
\definecolor{grey32}{rgb}{0.32,0.32,0.32}
\definecolor{grey33}{rgb}{0.33,0.33,0.33}
\definecolor{grey34}{rgb}{0.34,0.34,0.34}
\definecolor{grey35}{rgb}{0.35,0.35,0.35}
\definecolor{grey36}{rgb}{0.36,0.36,0.36}
\definecolor{grey37}{rgb}{0.37,0.37,0.37}
\definecolor{grey38}{rgb}{0.38,0.38,0.38}
\definecolor{grey39}{rgb}{0.39,0.39,0.39}
\definecolor{grey3}{rgb}{0.03,0.03,0.03}
\definecolor{grey40}{rgb}{0.40,0.40,0.40}
\definecolor{grey41}{rgb}{0.41,0.41,0.41}
\definecolor{grey42}{rgb}{0.42,0.42,0.42}
\definecolor{grey43}{rgb}{0.43,0.43,0.43}
\definecolor{grey44}{rgb}{0.44,0.44,0.44}
\definecolor{grey45}{rgb}{0.45,0.45,0.45}
\definecolor{grey46}{rgb}{0.46,0.46,0.46}
\definecolor{grey47}{rgb}{0.47,0.47,0.47}
\definecolor{grey48}{rgb}{0.48,0.48,0.48}
\definecolor{grey49}{rgb}{0.49,0.49,0.49}
\definecolor{grey4}{rgb}{0.04,0.04,0.04}
\definecolor{grey50}{rgb}{0.50,0.50,0.50}
\definecolor{grey51}{rgb}{0.51,0.51,0.51}
\definecolor{grey52}{rgb}{0.52,0.52,0.52}
\definecolor{grey53}{rgb}{0.53,0.53,0.53}
\definecolor{grey54}{rgb}{0.54,0.54,0.54}
\definecolor{grey55}{rgb}{0.55,0.55,0.55}
\definecolor{grey56}{rgb}{0.56,0.56,0.56}
\definecolor{grey57}{rgb}{0.57,0.57,0.57}
\definecolor{grey58}{rgb}{0.58,0.58,0.58}
\definecolor{grey59}{rgb}{0.59,0.59,0.59}
\definecolor{grey5}{rgb}{0.05,0.05,0.05}
\definecolor{grey60}{rgb}{0.60,0.60,0.60}
\definecolor{grey61}{rgb}{0.61,0.61,0.61}
\definecolor{grey62}{rgb}{0.62,0.62,0.62}
\definecolor{grey63}{rgb}{0.63,0.63,0.63}
\definecolor{grey64}{rgb}{0.64,0.64,0.64}
\definecolor{grey65}{rgb}{0.65,0.65,0.65}
\definecolor{grey66}{rgb}{0.66,0.66,0.66}
\definecolor{grey67}{rgb}{0.67,0.67,0.67}
\definecolor{grey68}{rgb}{0.68,0.68,0.68}
\definecolor{grey69}{rgb}{0.69,0.69,0.69}
\definecolor{grey6}{rgb}{0.06,0.06,0.06}
\definecolor{grey70}{rgb}{0.70,0.70,0.70}
\definecolor{grey71}{rgb}{0.71,0.71,0.71}
\definecolor{grey72}{rgb}{0.72,0.72,0.72}
\definecolor{grey73}{rgb}{0.73,0.73,0.73}
\definecolor{grey74}{rgb}{0.74,0.74,0.74}
\definecolor{grey75}{rgb}{0.75,0.75,0.75}
\definecolor{grey76}{rgb}{0.76,0.76,0.76}
\definecolor{grey77}{rgb}{0.77,0.77,0.77}
\definecolor{grey78}{rgb}{0.78,0.78,0.78}
\definecolor{grey79}{rgb}{0.79,0.79,0.79}
\definecolor{grey7}{rgb}{0.07,0.07,0.07}
\definecolor{grey80}{rgb}{0.80,0.80,0.80}
\definecolor{grey81}{rgb}{0.81,0.81,0.81}
\definecolor{grey82}{rgb}{0.82,0.82,0.82}
\definecolor{grey83}{rgb}{0.83,0.83,0.83}
\definecolor{grey84}{rgb}{0.84,0.84,0.84}
\definecolor{grey85}{rgb}{0.85,0.85,0.85}
\definecolor{grey86}{rgb}{0.86,0.86,0.86}
\definecolor{grey87}{rgb}{0.87,0.87,0.87}
\definecolor{grey88}{rgb}{0.88,0.88,0.88}
\definecolor{grey89}{rgb}{0.89,0.89,0.89}
\definecolor{grey8}{rgb}{0.08,0.08,0.08}
\definecolor{grey90}{rgb}{0.90,0.90,0.90}
\definecolor{grey91}{rgb}{0.91,0.91,0.91}
\definecolor{grey92}{rgb}{0.92,0.92,0.92}
\definecolor{grey93}{rgb}{0.93,0.93,0.93}
\definecolor{grey94}{rgb}{0.94,0.94,0.94}
\definecolor{grey95}{rgb}{0.95,0.95,0.95}
\definecolor{grey96}{rgb}{0.96,0.96,0.96}
\definecolor{grey97}{rgb}{0.97,0.97,0.97}
\definecolor{grey98}{rgb}{0.98,0.98,0.98}
\definecolor{grey99}{rgb}{0.99,0.99,0.99}
\definecolor{grey9}{rgb}{0.09,0.09,0.09}
\definecolor{grey}{rgb}{0.75,0.75,0.75}
\definecolor{honeydew1}{rgb}{0.94,1.00,0.94}
\definecolor{honeydew2}{rgb}{0.88,0.93,0.88}
\definecolor{honeydew3}{rgb}{0.76,0.80,0.76}
\definecolor{honeydew4}{rgb}{0.51,0.55,0.51}
\definecolor{honeydew}{rgb}{0.94,1.00,0.94}
\definecolor{hotpink}{rgb}{1.00,0.41,0.71}
\definecolor{indianred}{rgb}{0.80,0.36,0.36}
\definecolor{ivory1}{rgb}{1.00,1.00,0.94}
\definecolor{ivory2}{rgb}{0.93,0.93,0.88}
\definecolor{ivory3}{rgb}{0.80,0.80,0.76}
\definecolor{ivory4}{rgb}{0.55,0.55,0.51}
\definecolor{ivory}{rgb}{1.00,1.00,0.94}
\definecolor{khaki1}{rgb}{1.00,0.96,0.56}
\definecolor{khaki2}{rgb}{0.93,0.90,0.52}
\definecolor{khaki3}{rgb}{0.80,0.78,0.45}
\definecolor{khaki4}{rgb}{0.55,0.53,0.31}
\definecolor{khaki}{rgb}{0.94,0.90,0.55}
\definecolor{lavenderblush}{rgb}{1.00,0.94,0.96}
\definecolor{lavender}{rgb}{0.90,0.90,0.98}
\definecolor{lawngreen}{rgb}{0.49,0.99,0.00}
\definecolor{lemonchiffon}{rgb}{1.00,0.98,0.80}
\definecolor{lightblue}{rgb}{0.68,0.85,0.90}
\definecolor{lightcoral}{rgb}{0.94,0.50,0.50}
\definecolor{lightcyan}{rgb}{0.88,1.00,1.00}
\definecolor{lightgoldenrod}{rgb}{0.93,0.87,0.51}
\definecolor{lightgoldenrod}{rgb}{0.98,0.98,0.82}
\definecolor{lightgray}{rgb}{0.83,0.83,0.83}
\definecolor{lightgreen}{rgb}{0.56,0.93,0.56}
\definecolor{lightgrey}{rgb}{0.83,0.83,0.83}
\definecolor{lightpink}{rgb}{1.00,0.71,0.76}
\definecolor{lightsalmon}{rgb}{1.00,0.63,0.48}
\definecolor{lightsea}{rgb}{0.13,0.70,0.67}
\definecolor{lightsky}{rgb}{0.53,0.81,0.98}
\definecolor{lightslate}{rgb}{0.47,0.53,0.60}
\definecolor{lightslate}{rgb}{0.47,0.53,0.60}
\definecolor{lightslate}{rgb}{0.52,0.44,1.00}
\definecolor{lightsteel}{rgb}{0.69,0.77,0.87}
\definecolor{lightyellow}{rgb}{1.00,1.00,0.88}
\definecolor{limegreen}{rgb}{0.20,0.80,0.20}
\definecolor{linen}{rgb}{0.98,0.94,0.90}
\definecolor{magenta1}{rgb}{1.00,0.00,1.00}
\definecolor{magenta2}{rgb}{0.93,0.00,0.93}
\definecolor{magenta3}{rgb}{0.80,0.00,0.80}
\definecolor{magenta4}{rgb}{0.55,0.00,0.55}
\definecolor{magenta}{rgb}{1.00,0.00,1.00}
\definecolor{maroon1}{rgb}{1.00,0.20,0.70}
\definecolor{maroon2}{rgb}{0.93,0.19,0.65}
\definecolor{maroon3}{rgb}{0.80,0.16,0.56}
\definecolor{maroon4}{rgb}{0.55,0.11,0.38}
\definecolor{maroon}{rgb}{0.69,0.19,0.38}
\definecolor{mediumaquamarine}{rgb}{0.40,0.80,0.67}
\definecolor{mediumblue}{rgb}{0.00,0.00,0.80}
\definecolor{mediumorchid}{rgb}{0.73,0.33,0.83}
\definecolor{mediumpurple}{rgb}{0.58,0.44,0.86}
\definecolor{mediumsea}{rgb}{0.24,0.70,0.44}
\definecolor{mediumslate}{rgb}{0.48,0.41,0.93}
\definecolor{mediumspring}{rgb}{0.00,0.98,0.60}
\definecolor{mediumturquoise}{rgb}{0.28,0.82,0.80}
\definecolor{mediumviolet}{rgb}{0.78,0.08,0.52}
\definecolor{midnightblue}{rgb}{0.10,0.10,0.44}
\definecolor{mintcream}{rgb}{0.96,1.00,0.98}
\definecolor{mistyrose}{rgb}{1.00,0.89,0.88}
\definecolor{moccasin}{rgb}{1.00,0.89,0.71}
\definecolor{navajowhite}{rgb}{1.00,0.87,0.68}
\definecolor{navyblue}{rgb}{0.00,0.00,0.50}
\definecolor{navy}{rgb}{0.00,0.00,0.50}
\definecolor{oldlace}{rgb}{0.99,0.96,0.90}
\definecolor{olivedrab}{rgb}{0.42,0.56,0.14}
\definecolor{orange1}{rgb}{1.00,0.65,0.00}
\definecolor{orange2}{rgb}{0.93,0.60,0.00}
\definecolor{orange3}{rgb}{0.80,0.52,0.00}
\definecolor{orange4}{rgb}{0.55,0.35,0.00}
\definecolor{orangered}{rgb}{1.00,0.27,0.00}
\definecolor{orange}{rgb}{1.00,0.65,0.00}
\definecolor{orchid1}{rgb}{1.00,0.51,0.98}
\definecolor{orchid2}{rgb}{0.93,0.48,0.91}
\definecolor{orchid3}{rgb}{0.80,0.41,0.79}
\definecolor{orchid4}{rgb}{0.55,0.28,0.54}
\definecolor{orchid}{rgb}{0.85,0.44,0.84}
\definecolor{palegoldenrod}{rgb}{0.93,0.91,0.67}
\definecolor{palegreen}{rgb}{0.60,0.98,0.60}
\definecolor{paleturquoise}{rgb}{0.69,0.93,0.93}
\definecolor{paleviolet}{rgb}{0.86,0.44,0.58}
\definecolor{papayawhip}{rgb}{1.00,0.94,0.84}
\definecolor{peachpuff}{rgb}{1.00,0.85,0.73}
\definecolor{peru}{rgb}{0.80,0.52,0.25}
\definecolor{pink1}{rgb}{1.00,0.71,0.77}
\definecolor{pink2}{rgb}{0.93,0.66,0.72}
\definecolor{pink3}{rgb}{0.80,0.57,0.62}
\definecolor{pink4}{rgb}{0.55,0.39,0.42}
\definecolor{pink}{rgb}{1.00,0.75,0.80}
\definecolor{plum1}{rgb}{1.00,0.73,1.00}
\definecolor{plum2}{rgb}{0.93,0.68,0.93}
\definecolor{plum3}{rgb}{0.80,0.59,0.80}
\definecolor{plum4}{rgb}{0.55,0.40,0.55}
\definecolor{plum}{rgb}{0.87,0.63,0.87}
\definecolor{powderblue}{rgb}{0.69,0.88,0.90}
\definecolor{purple1}{rgb}{0.61,0.19,1.00}
\definecolor{purple2}{rgb}{0.57,0.17,0.93}
\definecolor{purple3}{rgb}{0.49,0.15,0.80}
\definecolor{purple4}{rgb}{0.33,0.10,0.55}
\definecolor{purple}{rgb}{0.63,0.13,0.94}
\definecolor{red1}{rgb}{1.00,0.00,0.00}
\definecolor{red2}{rgb}{0.93,0.00,0.00}
\definecolor{red3}{rgb}{0.80,0.00,0.00}
\definecolor{red4}{rgb}{0.55,0.00,0.00}
\definecolor{red}{rgb}{1.00,0.00,0.00}
\definecolor{rosybrown}{rgb}{0.74,0.56,0.56}
\definecolor{royalblue}{rgb}{0.25,0.41,0.88}
\definecolor{saddlebrown}{rgb}{0.55,0.27,0.07}
\definecolor{salmon1}{rgb}{1.00,0.55,0.41}
\definecolor{salmon2}{rgb}{0.93,0.51,0.38}
\definecolor{salmon3}{rgb}{0.80,0.44,0.33}
\definecolor{salmon4}{rgb}{0.55,0.30,0.22}
\definecolor{salmon}{rgb}{0.98,0.50,0.45}
\definecolor{sandybrown}{rgb}{0.96,0.64,0.38}
\definecolor{seagreen}{rgb}{0.18,0.55,0.34}
\definecolor{seashell1}{rgb}{1.00,0.96,0.93}
\definecolor{seashell2}{rgb}{0.93,0.90,0.87}
\definecolor{seashell3}{rgb}{0.80,0.77,0.75}
\definecolor{seashell4}{rgb}{0.55,0.53,0.51}
\definecolor{seashell}{rgb}{1.00,0.96,0.93}
\definecolor{sienna1}{rgb}{1.00,0.51,0.28}
\definecolor{sienna2}{rgb}{0.93,0.47,0.26}
\definecolor{sienna3}{rgb}{0.80,0.41,0.22}
\definecolor{sienna4}{rgb}{0.55,0.28,0.15}
\definecolor{sienna}{rgb}{0.63,0.32,0.18}
\definecolor{skyblue}{rgb}{0.53,0.81,0.92}
\definecolor{slateblue}{rgb}{0.42,0.35,0.80}
\definecolor{slategray}{rgb}{0.44,0.50,0.56}
\definecolor{slategrey}{rgb}{0.44,0.50,0.56}
\definecolor{snow1}{rgb}{1.00,0.98,0.98}
\definecolor{snow2}{rgb}{0.93,0.91,0.91}
\definecolor{snow3}{rgb}{0.80,0.79,0.79}
\definecolor{snow4}{rgb}{0.55,0.54,0.54}
\definecolor{snow}{rgb}{1.00,0.98,0.98}
\definecolor{springgreen}{rgb}{0.00,1.00,0.50}
\definecolor{steelblue}{rgb}{0.27,0.51,0.71}
\definecolor{tan1}{rgb}{1.00,0.65,0.31}
\definecolor{tan2}{rgb}{0.93,0.60,0.29}
\definecolor{tan3}{rgb}{0.80,0.52,0.25}
\definecolor{tan4}{rgb}{0.55,0.35,0.17}
\definecolor{tan}{rgb}{0.82,0.71,0.55}
\definecolor{thistle1}{rgb}{1.00,0.88,1.00}
\definecolor{thistle2}{rgb}{0.93,0.82,0.93}
\definecolor{thistle3}{rgb}{0.80,0.71,0.80}
\definecolor{thistle4}{rgb}{0.55,0.48,0.55}
\definecolor{thistle}{rgb}{0.85,0.75,0.85}
\definecolor{tomato1}{rgb}{1.00,0.39,0.28}
\definecolor{tomato2}{rgb}{0.93,0.36,0.26}
\definecolor{tomato3}{rgb}{0.80,0.31,0.22}
\definecolor{tomato4}{rgb}{0.55,0.21,0.15}
\definecolor{tomato}{rgb}{1.00,0.39,0.28}
\definecolor{turquoise1}{rgb}{0.00,0.96,1.00}
\definecolor{turquoise2}{rgb}{0.00,0.90,0.93}
\definecolor{turquoise3}{rgb}{0.00,0.77,0.80}
\definecolor{turquoise4}{rgb}{0.00,0.53,0.55}
\definecolor{turquoise}{rgb}{0.25,0.88,0.82}
\definecolor{violetred}{rgb}{0.82,0.13,0.56}
\definecolor{violet}{rgb}{0.93,0.51,0.93}
\definecolor{wheat1}{rgb}{1.00,0.91,0.73}
\definecolor{wheat2}{rgb}{0.93,0.85,0.68}
\definecolor{wheat3}{rgb}{0.80,0.73,0.59}
\definecolor{wheat4}{rgb}{0.55,0.49,0.40}
\definecolor{wheat}{rgb}{0.96,0.87,0.70}
\definecolor{whitesmoke}{rgb}{0.96,0.96,0.96}
\definecolor{white}{rgb}{1.00,1.00,1.00}
\definecolor{yellow1}{rgb}{1.00,1.00,0.00}
\definecolor{yellow2}{rgb}{0.93,0.93,0.00}
\definecolor{yellow3}{rgb}{0.80,0.80,0.00}
\definecolor{yellow4}{rgb}{0.55,0.55,0.00}
\definecolor{yellowgreen}{rgb}{0.60,0.80,0.20}
\definecolor{yellow}{rgb}{1.00,1.00,0.00}